\documentclass[twocolumn,amsmath,amssymb,superscriptaddress,10pt,aps]{revtex4-2}
\usepackage [utf8]{inputenc}
\usepackage{bm,physics}                        
\usepackage{graphicx}
\usepackage{xcolor}
\definecolor{brown}{rgb}{0.65,0.16,0.16} 
\usepackage[colorlinks,bookmarks=true,citecolor=blue,linkcolor=red,urlcolor=blue]{hyperref}
\usepackage[capitalise]{cleveref}
\crefname{section}{Appendix}{Appendices}
\usepackage{mathtools}

\newcommand{\vol}{\mathcal{V}}      
        
\newcommand{\MBZ}{\ensuremath{\mathrm{MBZ}}}

\DeclareMathOperator{\mbz}{MBZ}

\begin{document}
	\title{Umklapp-Enhanced Interlayer Valley Drag in Moiré Bilayers}
\author{Ritajit Kundu}
\affiliation{Institute of Mathematical Sciences, CIT Campus, Chennai 600113, India}
\author{Mandar M. Deshmukh}
\affiliation{Department of Condensed Matter Physics and Materials Science, Tata Institute of
Fundamental Research, Homi Bhabha Road, Mumbai 400005, India}
\author{Herbert A. Fertig}
\affiliation{Department of Physics, Indiana University, Bloomington, Indiana 47405, USA}
\affiliation{Quantum Science and Engineering Center, Indiana University, Bloomington, Indiana 47405, USA}
\author{Arijit Kundu}
\affiliation{Department of Physics, Indian Institute of Technology - Kanpur 208016, India}
	\begin{abstract}
Van der Waals materials may be combined to form moiré patterns that are effectively crystal lattices.  These systems are unique in that their in-plane unit cell sizes may be orders of magnitude larger than interlayer separations, leading to unique behaviors emerging from interlayer interactions. In this work, we investigate interlayer valley drag in lattice-matched moiré bilayers, demonstrating a remarkable enhancement due to umklapp scattering. In contrast to drag phenomena in more conventional two-dimensional systems, interlayer valley drag appears at first order in the interlayer interaction, and remains non-vanishing in the low temperature limit even at this low order in the interlayer coupling. We propose an experimental geometry, feasible with current state-of-the-art fabrication techniques, to detect and characterize this effect in moiré bilayer systems.
	\end{abstract}
\maketitle

\textit{Introduction} --
Interaction–induced drag phenomena provide a powerful window into many–body physics in low–dimensional electron systems.  In Coulomb drag experiments, an electrical current driven in an ``active'' layer generates, via interlayer interactions, a voltage or current in a nearby but electrically isolated ``passive'' layer~\cite{Rojo1999,Narozhny2016}. While initial work on GaAs quantum wells demonstrated the effectiveness of drag as a probe of two-dimensional electron gases (2DEGs)~\cite{Gramila1991,Sivan1992}, drag experiments have since uncovered rich physics in graphene bilayers~\cite{Kim2011,Gorbachev2012}, exciton condensates~\cite{Seamons2009,Nandi2012,Eisenstein_2014,PhysRevB.107.L041402}, and quantum Hall systems ~\cite{Kellogg2004,Tutuc2004}. 

Conventionally, when interlayer interaction is weak, drag is typically understood within perturbation theory in interlayer interactions $U_{12}$.  For Coulomb interactions, a characteristic result of this in time-reversal invariant systems is its vanishing at first order in $U_{12}$; moreover, the second order contribution typically vanishes quadratically with temperature $T$. The first of these arises from a precise cancellation between time-reversed scattering processes, while the second requires thermally excited particle-hole pairs to allow energy and momentum transfer between the two systems ~\cite{Kamenev1995,Flensberg1995,Song2013,Amorim2012}.  Valley degrees of freedom, in principle, open up further possibilities beyond these paradigms. Among these is interlayer valley drag, defined here as the generation of a valley current in one layer by the presence of such a current in a remote layer. This is expected to be suppressed in uniform systems, such as pristine graphene with the Fermi energy near its Dirac point, the valley drag susceptibility reduces to a Fermi-surface integral of the velocity that vanishes due to rotational symmetry.

In this work, we show that in moiré superlattices, this behavior is qualitatively changed. When two layers share a common moiré pattern, through careful alignment on opposite sides of a substrate crystal, or through a shared moiré potential \cite{Kim2024,Zeng_2026}, electrons in different layers may mutually scatter with momentum conserved only up to a moir\'e reciprocal lattice (mRL) $\bm{G}$. Because the moiré period $L_M$ is much larger than the interlayer separation $d$, these umklapp channels carry substantial interaction strength \cite{Jat_2024}. By contrast, for conventional 2DEGs, such processes are negligible in the drag response.
Importantly, non-vanishing valley drag appears at \textit{first} order in the interlayer interaction $U_{12}$, entirely from $\bm{G} \neq 0$ umklapp channels: the $\bm{G} = 0$ term vanishes identically, consistent with simpler systems such as pristine graphene at low-energy. Moreover, as we explain below, the behavior we find remains non-vanishing in the zero temperature limit.

The mechanism underlying this valley drag phenomenon may be understood qualitatively as follows. In each valley, a Fermi surface cuts through a band for which there is a specific microstructure associated with each unit cell.  A small ${\bf q}$ variation of the density can be implemented by a slowly varying Fermi energy, so that the (fractional) filling associated with that microstructure varies from cell to cell. Because there is structural matching between layers at the unit cell level, there is a strong response in the remote layer that effectively induces intercell potential differences, driving currents in each valley.  Time-reversal symmetry dictates that these currents are equal and opposite for valleys that are time-reversed partners, so that a charge current is absent; however, the remote layer develops a valley current.

While the basic valley drag phenomenon described here involves matched mRL's of the coupled systems, in principle, only possible for disorder-free systems, we also argue that minor misalignment does not eliminate the phenomenon.  Strain and structural disorder broaden the mRL peaks, as revealed by scanning tunneling microscopy studies of moiré systems~\cite{Uri2020,Kazmierczak2021}. The broadening relaxes strict inter-layer momentum-matching in the relevant scattering processes, so that layers with modest alignment disorder can exhibit finite valley drag, determined by the overlap of the broadened mRL peaks. 

Detection of valley drag requires coupling to valley-polarized currents, which are not directly accessible in standard transport. However, in systems with valley-contrasting Berry curvature, a longitudinal valley current generates a transverse charge accumulation, via the valley Hall effect~\cite{Xiao2007,Mak2014,Adak_2024}, which may be converted with a charge current with appropriately placed contacts. We propose an experimental geometry in which valley drag between aligned moiré bilayers is detected through nonlocal voltage measurements sensitive to this valley Hall response. Recent advances in the fabrication of dual-moiré heterostructures with independent control of the individual layer filling~\cite{Park2021,Cao2018a,Wang2019NanoLettDoublyAligned,wang2019composite,Sun2021NatCommunDoublyAligned,Hu2023NatCommunControlledAlignment,deVries2020PRLMinivalleyLayer,Wang2025PRLDoubleMoire} make such measurements feasible with current technology. In addition to its being of intrinsic physical interest, the observation of this phenomenon would offer new pathways to access valley transport in two-dimensional electron systems~\cite{Schaibley_2016,Xu_2025}.

\emph{Interlayer Valley Drag Conductivity}-- Consider two capacitively coupled two-dimensional electronic systems separated by a spacer. The layers $l=1,2$, each hosting two valleys [$\xi=(\mathrm K, \mathrm K')$] which are time-reversal partners. The valley-resolved current operators for a given direction $\alpha=\hat{x},\hat{y}$ is given by $j^{(\xi,\alpha)}_{l}$. 
In terms of these, the valley current operator in layer $l$ may be written as $j^{\alpha}_{l,v} = j^{(\mathrm{K},\alpha)}_{l}-j^{(\mathrm{K}',\alpha)}_{l}$ (see \cref{app:sec:ham}). 
To leading order in the interlayer interaction $U_{12}$, the valley drag conductivity (see \cref{app:sec:pert} for a derivation) can be written, separating intra- and inter-band contributions, as
	\begin{align}
		\sigma_{12,v}^{\alpha\beta} = -\frac{e^2}{h} \frac{2\pi}{\hbar \delta}
		\sum_{\bm G}
		&
		\mathcal C_{1v, - \bm G}^{\alpha} U_{12}(\bm G) \mathcal C_{2v,\bm G}^{\beta} \nonumber\\ &+ \text{inter-band terms},\label{eq:netsigma}
	\end{align}
    where we have assumed an energy-independent intralayer phenomenological disorder broadening parameterized by $\hbar \delta$. 
In metallic systems, the inter-band contribution is generally several orders of magnitude smaller than that of the intra-band processes (see \cref{app:sec:pert}), and thus is neglected in the following analysis. Here, $\mathcal C^{\alpha}_{lv,\bm G}$ is the valley-resolved current-density response functions (i.e., susceptibilities) from intra-band processes may be written as (see \cref{app:chi-deriv}) 
\begin{align}
	\mathcal C^{\alpha}_{lv,\bm G}
	&= \frac{2}{\hbar \vol}\sum_{n,\xi,\bm k} 
	\xi  f\big(\varepsilon_{l,n,\bm k,\xi}-\mu_l\big)\partial_{k_\alpha} 
    \lambda_{l,\bm G}^{(n,n,\xi)}(\bm k)
    \label{eq:chi}
\end{align}
where $\vol$ is the system area, $\varepsilon_{l,n,\bm k,\xi}$ is the energy of an electron with momentum ${\bm k}$ in the $n$th band, $f(\epsilon)$ is Fermi distribution function, and the form factor
\[
\lambda_{l,\bm G}^{(n,n,\xi)}(\bm k)
= \sum_{\bm G'}u^{\dagger }_{l,n,\bm k,\xi}(\bm G')u_{l,n,\bm k,\xi}(\bm G'{+}\bm G)
\]
involves the Fourier transform $u_{l,n\bm k,\xi}(\bm G)$ of the periodic part of Bloch states $u_{l,n,\bm k,\xi}(\bm r) = \sum_{\bm G} u_{l,n,\bm k,\xi}(\bm G) e^{i\bm G\cdot \bm r}$, with $\bm{G}$ being mRL vectors.
{Note that $\lambda_{l,\bm G}^{(n,n,\xi)}(\bm k)$ may be understood as the Fourier transform of the real space charge density associated with a Bloch state in a given layer and valley.} 
\begin{figure*}[htpb]
    \centering
    \includegraphics{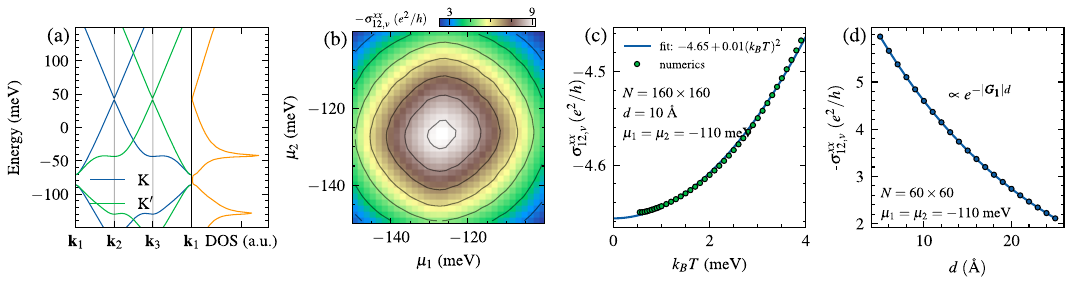}
    \caption{\textbf{Band-structure and valley drag condutivity}:
(a) Left: Band structure of aligned graphene/hBN heterostructure near the K and $\mathrm{K'}$ valleys of graphene, plotted along high-symmetry paths in the moir\'e Brillouin zone. 
For the K valley, the path connects $\mathbf{k}_1 = (1/2,\sqrt{3}/2)$, $\mathbf{k}_2 = (0,0)$, and $\mathbf{k}_3 = (1,0)$ and
for the $\mathrm{K'}$ valley, the path connects $\mathbf{k}_1 = (-1/2,\sqrt{3}/2)$, $\mathbf{k}_2 = (-1,0)$, and $\mathbf{k}_3 = (0,0)$, in units of $k_\epsilon$. 
Right: Density of states corresponding to these bands. 
(b) Valley drag conductivity $\sigma_{21,xx}^v$ as a function of the chemical potentials $\mu_1$ and $\mu_2$ of the top and bottom layers, respectively, computed for a system size $N = 60 \times 60$ at temperature $k_B T = 1\,\mathrm{meV}$. 
(c) Temperature dependence of $\sigma_{21,xx}^v$. 
(d) Dependence of $\sigma_{21,xx}^v$ on the interlayer separation $d$. 
We have used $\hbar \delta = 1$ meV.  
}
    \label{fig:result}
\end{figure*}

\textit{Current-Density Response}-- 
From \cref{eq:netsigma,eq:chi}, one sees that the form factors and the resulting current-density response functions play a crucial role in determining the valley drag of a system.  Several remarks on their properties are thus in order.
Firstly, we consider a situation in which the microstructure associated with a given band is at a small scale relative to the interlayer separation, in which case only the longest wavelength (${\bf G}=0$) term in Eq. \ref{eq:netsigma} has a $U_{12}$ differing significantly from zero.  The relevant form factor in this case obeys
$\lambda_{l,\bm 0}^{(n,n,\xi)}(\bm k)
= \sum_{\bm G'}u^{\dagger }_{l,n,\bm k,\xi}(\bm G')u_{l,n,\bm k,\xi}(\bm G') = 1$,  immediately
leading to $\mathcal C^{\alpha}_{lv,\bm 0}=0$. In this limit \textit{inter-band} contribution to the valley-drag also vanishes (see \cref{app:sec:pert}). Thus, the valley-drag conductivity vanishes.  
We note that this mechanism does not lead to charged current drag between the layers. At first order, the transconductivity has expressions very similar to those in \cref{eq:netsigma,eq:chi}, but with the valley index $\xi$ removed from the second. Time-reversal symmetry requires $v^{(n,\mathrm{K})}_{l,\alpha}(\bm k) = -v^{(n,\mathrm{K}')}_{l,\alpha}(-\bm k)$ and $\lambda_{l,\bm G}^{(n,n,\mathrm{K})}(\bm k) = [\lambda_{l,-\bm G}^{(n,n,\mathrm{K}')}(-\bm k)]^*$ (see \cref{app:TR}). Under the operation $(\mathrm{K},\bm k)\to(\mathrm{K}',-\bm k)$, the summand in the \emph{charge-drag susceptibility} (analog of Eq. \ref{eq:chi} without $\xi$ in the summand) changes sign, and so vanishes \textit{for any }$\bm G$.

Secondly, as stated above, for the systems we consider, the valley-drag phenomenon remains non-vanishing even as the temperature $T \to 0$.  To see that this can be the case, one integrates Eq. \ref{eq:chi} by parts, and notes that
at zero temperature $f'(\varepsilon{-}\mu)\to -\delta(\varepsilon{-}\mu)$.   This results in a Fermi-surface integral,
\begin{align}
	\mathcal C_{lv,\bm G}^{\alpha}(T=0)
	= \frac{2}{(2\pi)^2}\sum_{n,\xi}\oint_{\mathrm{FS}_{l,n,\xi}}
	&\frac{\xi d\ell_{\bm k}}{|{\bm v}_{l}^{(n,\xi)}|} v^{(n,\xi)}_{l,\alpha}
	\lambda^{(n,n,\xi)}_{l,\bm G},\label{eq:zerotemp}
\end{align}
where $\oint_{\mathrm{FS}_{l,n,\xi}}
d\ell_{\bm k}$ denotes an integral along the Fermi surface of band $n$ in layer $l$ and valley $\xi$, and the components of band velocity are $v^{(n,\xi)}_{l,\alpha} = \partial_{k_\alpha}\varepsilon_{l,n,\bm k,\xi}$. 
Again, one sees the result will vanish if only the ${\bm G=0}$ response were relevant; it is the presence of microstructure in the moiré unit cells at length scales for which the layers have significant coupling (i.e., umklapp processes) that allows a non-zero result to emerge at $T=0$.
This should be compared to conventional Coulomb drag, which arises at second order in $U_{12}$ and for which the phase space for interlayer scattering requires thermal excitation of particle-hole pairs in both layers, leading to a drag conductivity proportional to $T^2$ \cite{Narozhny2016,Levchenko2008}. Finally, at low but non-vanishing temperature, one may compute the responses relevant for the valley drag conductivity using the Sommerfeld expansion,
\begin{align*}
	\mathcal C_{lv,\bm G}^{\alpha}(T) \approx \mathcal C_{lv,\bm G}^{\alpha}(0) + \frac{\pi^2}{6}(k_BT)^2\frac{\partial^2 \mathcal C_{lv,\bm G}^{\alpha}(0)}{\partial \mu_l^2},
\end{align*}
so that one expects
$\sigma_{12,v}^{\alpha\beta}(T) - \sigma_{12,v}^{\alpha\beta}(T=0) \propto T^2 + \mathcal{O}(T^4).$
We verify this behavior in our numerical results, which we now turn to.

\emph{Numerical results for G/hBN/G}---
To illustrate the above results concretely, we consider a graphene/hBN/graphene (G/hBN/G) double-aligned structure, in which the $\sim\!1.8\%$ lattice mismatch at each graphene--hBN interface produces a moir\'e pattern with period $L_M \approx 14\,\mathrm{nm}$; the smallest reciprocal lattice vector of the resulting moir\'e lattice, $|\bm G_1| = 4\pi/(\sqrt{3}\, L_M)$, is approximately $0.52\,\mathrm{nm}^{-1}$. In addition to creating the moir\'e lattices, the insulating hBN layer serves as the interlayer spacer of thickness $d$, electrically isolating them while preserving their common mRL. The resulting band structure, shown in \cref{fig:result}(a), retains a Dirac cone inherited from graphene, supplemented by saddle points that produce prominent van Hove singularities (vHS) in the density of states.

\Cref{fig:result}(b) displays the valley drag conductivity $\sigma_{12,v}^{xx}$ as a function of the independently gate-tunable chemical potentials $\mu_1$ and $\mu_2$. At low temperatures, the response is dominated by the intraband Fermi-surface contribution of \cref{eq:netsigma}; the interband terms are quite small and have little effect, as discussed in more detail in the SM. The results exhibit the expected symmetry under layer interchange, $(\mu_1,\mu_2)\leftrightarrow(\mu_2,\mu_1)$, and peaks sharply when both chemical potentials approach a vHS, where the enhanced density of states amplifies the Fermi-surface integrals entering $\mathcal C^{\alpha}_{lv,\bm G}$.

Turning to the temperature dependence (\cref{fig:result}(c)), we verify the analytical prediction discussed above: away from the vHS (where the derivative in the Sommerfeld expansion is ill-defined), $\sigma_{12,v}^{xx}$ saturates to a finite value as $T\to 0$, with deviations that are well described by a $T^2$ correction.  As discussed above, this non-vanishing zero-temperature limit at first order in $U_{12}$ is in striking contrast to conventional charge Coulomb drag, where a zero temperature-contribution appears only at third order in the interlayer interaction~\cite{Levchenko2008}, making this behavior rather hard to observe when interlayer interactions are weak.

The dependence of the valley drag on spacer thickness $d$ (\cref{fig:result}(d)) provides a further consistency check: since the interlayer interaction at mRL vector $\bm G$ is suppressed by $e^{-|\bm G|\, d}$, the dominant umklapp contribution comes from the first shell of reciprocal lattice vectors, and the valley drag conductivity decays as $\sigma_{12,v}^{xx}\sim e^{-|\bm G_1|\, d}$. Because $|\bm G_1|$ is set by the large moir\'e period rather than the atomic lattice, this decay is slow, for $d = 10\,\text{\AA}$ one has $|\bm G_1| d \approx 0.5$, so the coupling remains substantial even for realistic hBN spacer thicknesses.  This is marked contrast to expectations for conventional crystals, for which, typically, $|{\bf G}_1|d \gg 1$, making the valley drag effect negligible in such systems.

\begin{figure}[htb]
	\centering
	\includegraphics{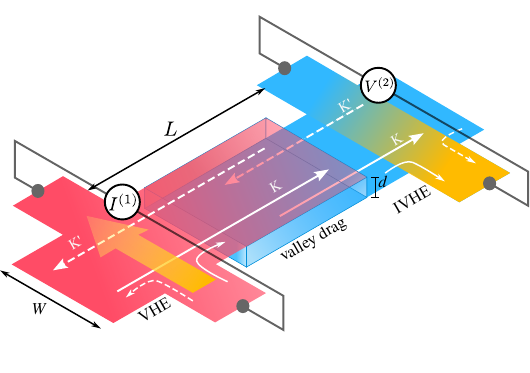}
	\caption{\textbf{Proposed experiment:} A current injected in the top (red) layer generates a transverse valley current via the valley Hall effect (VHE). The diffusing valley current drags a valley current in the bottom (blue) layer and induces a nonlocal voltage at a distance $L$ via the inverse valley Hall effect (IVHE). The layers have width $W$ and are separated by an hBN spacer of thickness $d$.}
	\label{fig:drag-setup}
\end{figure}

\emph{Observing Interlayer Valley Drag }---
A central challenge in detecting valley drag is that valley currents carry no net charge and therefore produce no voltage in a standard transport measurement. To overcome this, one may exploit the valley Hall effect, which in systems with valley-contrasting Berry curvature allows charge transport to induce valley currents~\cite{Xiao2007,Mak2014,Gorbachev2014}. This enables a three-stage detection scheme (Fig.~\ref{fig:drag-setup}), as follows. (i) A charge current in the drive layer induces a transverse valley current via the valley Hall effect. (ii) A region in which the valley current is present in one layer is overlaid above a second layer with a matching mRL.  Interlayer valley drag thus transfers the (active) valley current to the remote (passive) layer. (iii) The inverse valley Hall effect~\cite{Sui2015,Shimazaki2015}
In the passive layer, it converts the induced valley current back into a measurable charge current or voltage. This proposed device architecture is a generalization of single-layer non-local resistance measurement for measuring bulk valley transport \cite{Gorbachev2012,Beconcini_2016,Sinha_2020} to interlayer valley drag between two moir\'e systems.

Focusing on the proposed experimental setup shown in \cref{fig:drag-setup}, a charge current $I^{(1)} \sim \sigma^{xx}_{1} E_1$ driven through the top layer with charge conductivity $\sigma_{1}^{xx}$ and electric field $E_1$ generates a transverse valley current through the valley Hall conductivity $\sigma_{1,v}^{xy}$. This valley current is $j_{1,v}^{x} = \sigma^{xy}_{1,v}\, E_1$, and consists of oppositely propagating valley modes. In the G/hBN moir\'e system, the inversion symmetry is broken and ensures that $\sigma^{xy}_{1,v}$ is nonzero.
Interlayer valley drag then induces a corresponding valley current in the bottom layer,
$j^x_{2,v} \sim \sigma_{12,v}^{xx} j^x_{1,v} / \sigma_1^{xx}$.
In the absence of current leads in the passive layer, the inverse valley Hall effect converts this valley current into a nonlocal voltage drop,
$V^{(2)} \sim \sigma_{2,v}^{xy}j_{2,v}^x/(\sigma^{xx}_{2})^2 $,
where $\sigma_{2,v}^{xy}$ is the valley Hall conductivity of the bottom layer. Combining these relations, the resulting nonlocal resistance is
\begin{align}
	\frac{V^{(2)}}{I^{(1)}} \sim
	\left(\frac{L}{W}\right)
    \frac{\sigma^{xx}_{12,v}}{\sigma_1^{xx}\sigma_2^{xx}}
    \left(\frac{\sigma^{xy}_{1,v}}{{\sigma^{xx}_{1}}}\right)
    \left(\frac{\sigma^{xy}_{2,v}}{{\sigma^{xx}_{2}}}\right)
    e^{-L/l_v},
\end{align}
where $L$ is the probe separation and $W$ is the width of the system. The exponentially decaying part is for the diffusion of valley current away from the current source to the voltage probe, $l_v$ is the valley diffusion length. Previous experiment has reported $l_v \approx 1\,\mu$m \cite{Gorbachev2012}.

This geometry offers several signatures that distinguish it. For example, the nonlocal voltage $V^{(b)}$ changes sign under reversal of the drive current, $I^{(1)}\to -I^{(1)}$.  Moreover, it changes sign when the valley Hall response of either layer ($\sigma^{xy}_{1,v}$ or $\sigma^{xy}_{2,v}$) is reversed. Such changes could be accomplished, for example, by reversing a displacement field, which can change the sign of the Berry curvature in a moir\'e band \cite{Koshino_2019,Adak_2022}.Most importantly, the signal is sensitive to the relative alignment of the mRLs of the two layers: misalignment suppresses the umklapp form factors and hence the valley drag conductivity. Together with the predicted saturation of $\sigma^{xx}_{12,v}$ to a finite value as $T\to 0$, these properties provide a set of experimentally accessible signatures associated with the valley Hall effect.  

The interlayer valley Hall effect, as we have described it to this point, assumes lattices aligned perfectly with their substrate as well as with one another. In practice, this is difficult to achieve experimentally. However, moir\'e systems are prone to strain \cite{Huang_2018,Yoo_2019,de_beule_aharonov-bohm_2020,PhysRevB.110.085422,Turkel_2022,Devakul_2023,Mukherjee_2025,Xia_2025} and substrate disorder. These effects broaden the peaks in the Fourier transform of STM topographic images of moiré systems \cite{Kim_2023,Nuckolls_2023} and relax the strict momentum conservation condition. As a result, interlayer scattering will conserve momentum only to within the width of broadened peaks associated with the mRLs. In practice, we expect the valley Hall effect to be observable provided the broadened peaks of the two mRLs significantly overlap. While a microscopic treatment of this broadening is challenging, an estimate of its effect may be found phenomenologically, by broadening the correlation functions in \cref{eq:netsigma} to capture disorder effects. Concretely, if one replaces the discrete sum over reciprocal lattice vectors by a continuous distribution with Gaussian width $\eta$, assuming the broadened Bragg peaks of the two layers are aligned on average, we find that the correction to the drag conductivity scales as $\delta \sigma^{xx}_{12,v}\sim -\eta^{2}$ (see SM for details). For example, in magic-angle twisted trilayer graphene $\eta \approx 5-6$ nm$^{-1}$ \cite{Kim_2023}. 
Note that if the layers are so misaligned that their broadened peaks do not overlap, the valley drag conductivity vanishes at first order; the leading non-vanishing contribution is then expected at second order in the interlayer interaction strength, analogously to more standard results for interlayer drag. A related platform that naturally avoids such misalignment has recently been proposed in a moiré heterostructure where a twisted layer of hBN is sandwiched between two transition metal dichalcogenide (TMD) layers. In this system, the twisted hBN develops ferroelectric domains that generate a moiré potential in the adjacent TMD layers \cite{Kim2024,Zeng_2026}. Because this potential originates from the same hBN layer, the resulting moiré patterns in the two TMD layers are automatically aligned.

\textit{Conclusion} -- Moir\'e materials have already been shown to host a wide range of remarkable phenomena \cite{yankowitz2019tuning,Xie_2021,lu2019superconductors,Polshyn_2019,liu2021tuning,xie2019spectroscopic,choi2019electronic,nuckolls2020strongly,saito2021hofstadter,das2021symmetry,wu2021chern,potasz2021exact,kang2020non,hejazi2021hybrid,Dubey_2025}. In this work, we identify a new type of behavior: interlayer valley drag between aligned moir\'e lattices. In strong contrast to conventional drag mechanisms, this effect arises already at first order in the interlayer interaction $U_{12}$ due to umklapp processes, and it remains finite in the zero-temperature limit.  In more conventional materials, drag phenomena support a finite zero-temperature contribution only at third order in $U_{12}$, leading to a very small transconductance that has not yet been unambiguously observed~\cite{Levchenko2008,Narozhny2016}. Although this higher-order effect has not been observed in semiconductor samples, it has been proposed as a possible explanation for the finite drag resistivity measured in charge-neutral graphene \cite{Gorbachev2012}. 
The enhanced drag in matched moir\'e systems originates from their large moir\'e unit cells, which enable strong interlayer umklapp processes that are negligible in conventional 2DEGs. The effect should be observable in systems where pure valley currents can be generated via the valley Hall effect, particularly in moir\'e bands with nonvanishing Chern numbers.

\textit{Acknowledgements}--
R.K. acknowledges support from the FARE program at IIT Kanpur and the NPDF scheme of ANRF Grant No. PDF/2025/002176. M.M.D. acknowledges support from J.C. Bose Fellowship JCB/2022/000045 from the Department of Science and 32/33 Technology of India, and Department of Atomic Energy of the Government of India under award number 12-R\&D-TFR-5.10- 0100. H.A.F. acknowledges the support from the National Science Foundation Grant No. DMR-2531425.

\appendix

\begin{table*}[t]
\caption{Symbols used in this work and their descriptions}
\label{tab:params}
\begin{ruledtabular}
\begin{tabular}{ccc}
Symbol & Description \\
\hline
$\xi = (\mathrm K, \mathrm K') = (+1,-1)$ & valley index \\
$l = (1,2)$ & layer index \\
$n,m$ & band index \\
$\bm k, \bm q$ & Bloch momenta in the moir\'e Brillouin zone \\
$\bm G$ & moir\'e reciprocal lattice vector \\
$u_{n,\bm k,\xi}(\bm G)$ & Fourier component of periodic part of Bloch state \\
$\varepsilon_{n,\bm k,\xi}$ & Bloch band \\
$\alpha,\beta = (\hat x, \hat y)$ & index for spatial directions \\
Subscript-$s$ & $s=c$ for charge index and $s=v$ for valley index \\
$j_{ls}^{\alpha}$ & current operator \\
$\Gamma^{(\xi)}_{l;\alpha}$ & velocity matrix \\
$\Lambda_{ls;\alpha,\bm G}^{(n,m;\xi)}(\bm k,\bm q)$, $\lambda_{ls;\bm G}^{(n,m;\xi)}(\bm k,\bm q)$ & 
Current and density matrix element \\
$\Lambda_{ls;\alpha}^{(n,m;\xi)}(\bm k)$ & 
$\equiv \Lambda_{ls;\alpha,\bm 0}^{(n,m;\xi)}(\bm k,\bm 0)$ \\
$\lambda_{l;\bm G}^{(n,m;\xi)}(\bm k)$ &
$\equiv \lambda_{ls;\bm G}^{(n,m;\xi)}(\bm k,\bm 0)$ \\
$v^{(n,\xi)}_{l,\alpha}(\bm k)$ &
$=\Lambda^{(n,n,\xi)}_{lc;\alpha}(\bm k)$, band velocity \\
$\mathcal C^{\alpha}_{ls,\bm G}$,  $\mathcal D^{\alpha}_{ls,\bm G}$ & Intraband and interband current-density correlation function  \\
$H_{12}$ & Interlayer interaction \\
$\sigma_{12,s}^{\alpha \beta}$
&Interlayer drag conductivity\\
\end{tabular}
\end{ruledtabular}
\end{table*}

\section{Moiré Hamiltonian and parameters for graphene--hBN}
\label{app:sec:ham}

For explicit calculations of the valley-drag response we consider the moiré system formed in a graphene--hBN (G--hBN) heterostructure. In this system the moiré potential originates primarily from the lattice mismatch between graphene and hBN. Unlike twisted graphene systems, G--hBN does not host strong correlation-driven phases, which simplifies the calculation of transport responses. We employ the low-energy continuum Hamiltonian derived by Moon \textit{et al.}~\cite{Moon_2014}.

The G--hBN moiré system is described by the graphene Hamiltonian $H_{g,\xi}$ at valley $\xi = (\mathrm K, \mathrm K')$ coupled to the hBN Hamiltonian $H_{\mathrm{hBN}}$ through a spatially modulated moiré potential $U_\xi(\bm r)$. The full Hamiltonian reads
\begin{align}
H = \bigoplus_{\xi \in \{\mathrm K,\mathrm K'\}}
\begin{pmatrix}
H_{g,\xi} & U_\xi^\dagger(\bm r) \\
U_\xi(\bm r) & H_{\mathrm{hBN}}
\end{pmatrix},
\label{eq:g-hBN}
\end{align}
where
\begin{align}
U_\xi(\bm r) =
u_0\Bigg[
\begin{pmatrix}
1 & 1 \\ 1 & 1
\end{pmatrix}
+
\begin{pmatrix}
1 & \omega^* \\ \omega & 1
\end{pmatrix}
e^{i\xi \bm G_1^{\mathrm M}\cdot\bm r}
 \nonumber \\ 
 +
\begin{pmatrix}
1 & \omega \\ \omega^* & 1
\end{pmatrix}
e^{i\xi \bm G_2^{\mathrm M}\cdot\bm r}
\Bigg],
\end{align}
with $\omega = e^{2\pi i/3}$. The graphene Dirac Hamiltonian at valley $\xi$ is
\begin{align}
H_{g,\xi} = \hbar v_f \left(\xi p_x \sigma_x + p_y \sigma_y \right),
\end{align}
while the hBN layer is approximated by
\begin{align}
H_{\mathrm{hBN}} \approx
\begin{pmatrix}
V_N & 0 \\
0 & V_B
\end{pmatrix}.
\end{align}
The Hamiltonian in \cref{eq:g-hBN} acts on the four-component spinor $(\psi_1,\psi_2,\tilde\psi_1,\tilde\psi_2)$, where $\psi_1$ and $\psi_2$ denote the two graphene sublattices, while $\tilde\psi_1$ and $\tilde\psi_2$ correspond to the sublattices of the hBN layer.
The moiré reciprocal lattice vectors are
\(
\bm G_1^{\mathrm M} = k_m \left(\frac{3}{2},-\frac{\sqrt{3}}{2}\right)\), \(
\bm G_2^{\mathrm M} = \sqrt{3}k_m (0,1)
\),
where the moiré momentum scale is
\(
k_m = \frac{4\pi \epsilon}{3a(1+\epsilon)}\),
with lattice mismatch $\epsilon = a_{\mathrm{hBN}}/a - 1$. The lattice constants are $a_{\mathrm{hBN}} = 2.504$~\AA\ and $a = 2.46$~\AA\ for hBN and graphene, respectively. We use a Fermi velocity $v_f = 0.8\times10^6$ m/s and a moiré potential strength $u_0 = 152$ meV. The hBN onsite energies are taken as $V_N = 3.35$ eV and $V_B = -1.40$ eV~\cite{Moon_2014}.

The Bloch bands are obtained by diagonalizing Eq.~(\ref{eq:g-hBN}) in a plane-wave basis. We include moiré reciprocal lattice vectors
\(
\bm G_{nm} = n\bm G_1^{\mathrm M} + m\bm G_2^{\mathrm M}
\)
with $\abs{n}\le3$ and $\abs{m}\le3$. Increasing this cutoff does not significantly modify the low-energy bands or Bloch states, indicating convergence of the calculation. The resulting band dispersion and density of states used in the main text [Fig.~\ref{fig:result}(a)] are computed from this Hamiltonian.

In the eigenbasis of \cref{eq:g-hBN}, the Hamiltonian takes the form
\begin{align}
H = \sum_{n,\bm k,\xi,\sigma}
\epsilon_{n\bm k\xi}
c^{\dagger}_{n\bm k\xi\sigma}
c_{n\bm k\xi\sigma},
\end{align}
where $n$ denotes the Bloch-band index and $\bm k \in \MBZ$ lies within the moiré Brillouin zone.

The Bloch eigenstates are
\begin{align}
&\phi_{n,\bm k,\xi,\sigma}(\bm x)
= e^{i(\bm k+\bm K_{\xi})\cdot\bm x}\,
u_{n,\bm k,\xi,\sigma}(\bm x)
\\
&=
\frac{1}{\sqrt{A_{\mathrm{uc}}}}
\sum_{\bm G}
e^{i(\bm k+\bm G+\bm K_{\xi})\cdot\bm x}
\, v_{\sigma} \otimes u_{\xi} \otimes
u_{n,\bm k,\xi}(\bm G).
\label{eq:blochstate}
\end{align}
In the second line we have expanded the cell-periodic function
$u_{n,\bm k,\xi,\sigma}(\bm x)$ in a Fourier series over moiré reciprocal
lattice vectors $\bm G$ with Fourier coefficients $u_{n,\bm k,\xi}(\bm G)$. Here $A_{\mathrm{uc}}$ is the moiré unit-cell area.

Spin and valley are good quantum numbers. Their basis vectors are
$u_{\uparrow}=(1,0)^t$, $u_{\downarrow}=(0,1)^t$ for spin and
$v_{\mathrm K}=(1,0)^t$, $v_{\mathrm{K'}}=(0,1)^t$ for valley.
We denote the valley index as $\xi=(\mathrm K,\mathrm{K}')=(+1,-1)$
and the spin index as $\sigma=(\uparrow,\downarrow)$.
The valley momenta are $\bm K_{\xi}= \frac{4\pi}{3a}(\xi,0)$.

The field operator is
\begin{align}
\Psi(\bm x)
=
\frac{1}{\sqrt N}
\sum_{n,\bm k,\xi,\sigma}
\phi_{n,\bm k,\xi,\sigma}(\bm x)
c_{n\bm k\xi\sigma},
\end{align}
where $N$ is the number of moiré unit cells in the system.
The total area of the system is $\vol = N A_{\mathrm{uc}}$.

\begin{widetext}
The density operator $\rho(\bm x)=\Psi^\dagger(\bm x)\Psi(\bm x)$ can be written as
\begin{align}
\rho(\bm x)
&=
\frac{1}{\vol}
\sum_{\bm q,\bm G}
e^{i(\bm q+\bm G)\cdot\bm x}
\rho(\bm q+\bm G),
\\
\rho(\bm q+\bm G)
&=
\sum_{n_1,n_2,\xi,\sigma}
\sum_{\bm k}
\lambda^{(n_1,n_2;\xi)}_{\bm G}(\bm k,\bm q)
c^\dagger_{n_1,\bm k,\xi,\sigma}
c_{n_2,\mbz(\bm k+\bm q),\xi,\sigma}.
\end{align}

The density form factor is
\begin{align}
\lambda^{(n,m,\xi)}_{\bm G}(\bm k,\bm q)
=
\sum_{\bm G_1}
u^\dagger_{n,\bm k,\xi}(\bm G_1)
u_{m,\mbz(\bm k+\bm q),\xi}
(\bm G_1+\bm G+\bm G_{\bm k+\bm q}).
\label{eq:form-fac-i}
\end{align}

The momentum $\bm k+\bm q$ is decomposed into a reciprocal lattice vector
and a momentum within the moiré Brillouin zone,
$\bm k+\bm q=\mbz(\bm k+\bm q)+\bm G_{\bm k+\bm q}$,
where $\bm G_{\bm k+\bm q}$ is the quotient and
$\mbz(\bm k+\bm q)$ is the remainder lying inside the \MBZ.

The current operator is written as
\begin{align}
j_s^{\alpha}(\bm x)
&=
\frac{1}{\vol}
\sum_{\bm q,\bm G}
e^{i(\bm q+\bm G)\cdot\bm x}
j_s^{\alpha}(\bm q+\bm G),
\\
j_s^{\alpha}(\bm q+\bm G)
&=
\sum_{n_1,n_2,\xi,\sigma,\bm k}
\Lambda_{s;\alpha,\bm G}^{(n_1,n_2,\xi)}(\bm k,\bm q)
c^\dagger_{n_1,\bm k,\xi,\sigma}
c_{n_2,\mbz(\bm k+\bm q),\xi,\sigma},
\end{align}
where $s=(c,v)$ denotes charge 
The corresponding current form factors are
\begin{align}
\Lambda_{c;\alpha,\bm G}^{(n,m,\xi)}(\bm k,\bm q)
&=
\sum_{\bm G_1}
u^\dagger_{n,\bm k,\xi}(\bm G_1)
\Gamma^{(\xi)}_{\alpha}
u_{m,\mbz(\bm k+\bm q),\xi}
(\bm G_1+\bm G+\bm G_{\bm k+\bm q}),
\label{eq:form-fac-ii}
\\
\Lambda_{v;\alpha,\bm G}^{(n,m,\xi)}(\bm k,\bm q)
&=
\sum_{\bm G_1}
\xi\,
u^\dagger_{n,\bm k,\xi}(\bm G_1)
\Gamma^{(\xi)}_{\alpha}
u_{m,\mbz(\bm k+\bm q),\xi}
(\bm G_1+\bm G+\bm G_{\bm k+\bm q}).
\label{eq:form-fac-iii}
\end{align}

Here $\Gamma^{(\xi)}_{\alpha}=\frac{1}{\hbar}\frac{\partial H^{(\xi)}(\bm k)}{\partial k_\alpha}$ is the velocity matrix. For the G--hBN system, the current operator is the direct sum of the graphene and hBN contributions. Since the low-energy model of hBN is momentum independent, it does not contribute to the current. Denoting a $2\times2$ zero matrix by $\bm 0_2$, the velocity matrices are
$\Gamma^{(\xi)}_x = e v_f \xi \sigma_x \oplus \bm 0_2$ and
$\Gamma^{(\xi)}_y = v_f \sigma_y \oplus \bm 0_2$,
where $\sigma_x$ and $\sigma_y$ are Pauli matrices.

So far, we have discussed a single G--hBN heterostructure. In the next section, we consider two such layers and compute the interlayer drag response. At that stage, we introduce an additional layer index $l$ to distinguish the spectral properties of the top and bottom layers.
\end{widetext}

\section{Perturbation theory for drag conductivity}
\label{app:sec:pert}
Following \cite{Kamenev1995,Flensberg1995}, we consider two electronic systems coupled purely through capacitive interactions. The total Hamiltonian is
\begin{align}
    H = H_1 + H_2 + H_{12},
\end{align}
where $H_1$ and $H_2$ describe two spatially separated (by a distance $d$) but otherwise identical G--hBN heterostructures, each governed by the single-layer Hamiltonian in \cref{eq:g-hBN}. In general, $H_1$ and $H_2$ may contain disorder and interaction terms in addition to the non–interacting band structure.

Experimentally, such a system can be realized by placing graphene layers on the top and bottom of a few-layer-thick hBN spacer. Both graphene layers are aligned with the hBN, producing moiré superlattices in each layer. The hBN spacer simultaneously suppresses interlayer tunneling while allowing long-range Coulomb interactions.

The interlayer coupling is described by the density–density interaction
\begin{align}
    H_{12} &= \int \dd \bm r \, \dd \bm r'\;
    \rho_1(\bm r)\, U_{12}(\bm r-\bm r')\, \rho_2(\bm r')  \\
    &= \frac{1}{\vol}
    \sum_{\bm q \in \MBZ}\sum_{\bm G}
    U_{12}(\bm q+\bm G)\,
    \rho_1(-\bm q-\bm G)\,
    \rho_2(\bm q+\bm G).
\end{align}
Here $U_{12}$ is the Coulomb interaction between the two layers and $\rho_i$ is the density operator of layer $i$. The momentum $\bm q$ lies within the moiré Brillouin zone (\MBZ), while $\bm G$ denotes a moiré reciprocal lattice vector. Throughout this section, we adopt the convention that bold lowercase symbols (e.g.\ $\bm q,\bm k$) denote momenta within the \MBZ, while bold uppercase symbols (e.g.\ $\bm G,\bm G_1$) denote moiré reciprocal lattice vectors.

The DC drag conductivity is obtained from the interlayer current–current correlation function,
\begin{align}
\sigma_{12,s}^{\alpha \beta}
=
- \lim_{\Omega \to 0}
\lim_{\substack{\bm q \to 0\\ \bm G \to 0\\ \bm G' \to 0}}
\frac{1}{\Omega}
\Im \Big[
\Pi_{12,s}^{\alpha \beta}(\bm q+\bm G,-\bm q-\bm G',\Omega)
\Big].
\end{align}
Here $\alpha,\beta=(\hat x,\hat y)$ label spatial directions and $s=(c,v)$ denotes charge or valley response, respectively. The quantity $\Pi_{12,s}^{\alpha\beta}$ is the Fourier component of the interlayer current–current correlation function defined below.

Treating $H_{12}$ as a perturbation to $H_1+H_2$, the correlation function can be evaluated using the standard $S$–matrix expansion. In imaginary time the correlation function reads
\begin{align}
\Pi_{12,s}^{\alpha\beta}(\bm x\tau,\bm x'\tau')
&=
- \expval{T_\tau
j_{1s}^\alpha(\bm x\tau)\,
j_{2s}^\beta(\bm x'\tau')}
\\
&=
-
\frac{
\expval{
T_\tau
S(\beta)
\hat j_{1s}^\alpha(\bm x\tau)
\hat j_{2s}^\beta(\bm x'\tau')
}
}{
\expval{S(\beta)}
},
\label{eq:corr}
\end{align}
where in the second line we have moved to the interaction picture. Operators with carets evolve by the individual layer Hamiltonians. The operator $j_{is}^\alpha$ represents the charge or valley current of layer $i$ in $\alpha$ direction.

The $S$–matrix is expanded perturbatively in $H_{12}$,
\begin{align}
S(\beta)
&=
T_\tau
\exp\!\left(
-
\int_0^\beta \dd\tau_1\,
\hat H_{12}(\tau_1)
\right)
\\
&\approx
1-
\int_0^\beta \dd\tau_1\,
\hat H_{12}(\tau_1),
\label{eq:S-matrix}
\end{align}
where $\beta = 1/k_B T$ and $T_\tau$ denotes imaginary-time ordering.

Substituting \cref{eq:S-matrix} into \cref{eq:corr} allows the correlation function to be computed order by order in the interlayer interaction $H_{12}$. The zeroth-order contribution vanishes since, in the absence of $H_{12}$, the current operators in the two layers are completely decoupled.

\begin{widetext}
The first-order contribution to the correlation function is given by
\begin{align}
\Pi_{12,s}^{\alpha \beta}(\bm x \tau, \bm x' \tau')
=
\frac{1}{\vol}\sum_{\bm q \in \MBZ, \bm G}
\int_0^\beta \dd \tau_1
\expval{
\hat j_{1s}^\alpha(\bm x \tau)
\hat \rho_1(\bm q+\bm G,\tau_1)
}
U_{12}(\bm q + \bm G)
\expval{
\hat \rho_2(-\bm q-\bm G,\tau_1)
\hat j_{2s}^\beta(\bm x' \tau')
},
\label{eq:first-order}
\end{align}
where the expectation values are taken with respect to the ground states of the uncoupled individual layers. Fourier transforming the correlation function yields
\begin{align}
\Pi_{12,s}^{\alpha \beta}(\bm q + \bm G_1, - \bm q - \bm G_2, i\Omega_n)
=
\sum_{\bm G}
\chi_{1s}^{\alpha}(\bm q + \bm G_1, -\bm q - \bm G, i \Omega_n)
U_{12}(\bm q + \bm G)
\chi_{2s}^{\beta}(\bm q + \bm G, - \bm q - \bm G_2, i \Omega_n),
\end{align}
where $\Omega_n$ is a bosonic Matsubara frequency. In arriving at this expression, the expectation values in \cref{eq:first-order} factorize because the two layers are decoupled in the absence of $H_{12}$. For simplicity, we further assume that each layer is non–interacting.
The Fourier components of the individual response functions are then
\begin{align}
\chi_{1s}^{\alpha}(\bm q + \bm G_1, -\bm q - \bm G, i \Omega_n)
&=
\frac{2}{\vol}
\sum_{n,m,\bm k,\xi}
\Lambda^{(n,m,\xi)}_{1s;\alpha \bm G_1}(\bm k, \bm q)
\lambda^{(m,n,\xi)}_{1;-\bm G}(\bm k + \bm q, -\bm q)
\frac{
f(\varepsilon_{1,n,\bm k,\xi} - \mu_{1}) -
f(\varepsilon_{1,m,\bm k + \bm q,\xi} - \mu_{1})
}{
i \Omega_n + \varepsilon_{1,n,\bm k,\xi} - \varepsilon_{1,m,\bm k + \bm q,\xi}
},
\\
\chi_{2s}^{\beta}(\bm q + \bm G, - \bm q - \bm G_2, i \Omega_n)
&=
\frac{2}{\vol}
\sum_{n,m,\bm k,\xi}
\lambda^{(n,m,\xi)}_{2;\bm G}(\bm k , \bm q)
\Lambda^{(m,n,\xi)}_{2s;\beta -\bm G_2}(\bm k + \bm q, -\bm q)
\frac{
f(\varepsilon_{2,n,\bm k,\xi} - \mu_{2}) -
f(\varepsilon_{2,m,\bm k + \bm q,\xi} - \mu_{2})
}{
i \Omega_n + \varepsilon_{2,n,\bm k,\xi} - \varepsilon_{2,m,\bm k + \bm q,\xi}
}.
\end{align}

The prefactor of $2$ arises from spin degeneracy. Here $f(z) = (\exp(\beta z) + 1)^{-1}$ is the Fermi distribution function, $\varepsilon_{l,n,\bm k,\xi}$ are the Bloch-band energies obtained from the eigenvalues of \cref{eq:g-hBN}, and $\mu_1,\mu_2$ are the chemical potentials of the two layers. The quantities $\Lambda_{ls;\alpha,\bm G}^{(n,m,\xi)}(\bm k,\bm q)$ and $\lambda_{l;\bm G}^{(n,m,\xi)}(\bm k,\bm q)$ are, respectively, current and density form factors between Bloch bands $n,m$ in valley $\xi$ of layer $l$ as given in \cref{eq:form-fac-i,eq:form-fac-ii,eq:form-fac-iii}.

We first take the limit $\bm q,\bm G_1,\bm G_2 \to 0$, corresponding to a spatially uniform external perturbation. Next, we analytically continue the Matsubara frequency according to $i\Omega_n \to \Omega + i\delta^+$, divide the correlation function by $\Omega$, and finally take the limits $\Omega \to 0$ and $\delta \to 0$. After a straightforward but lengthy calculation, the DC drag conductivity to first order is obtained as
\begin{align}
\sigma_{12,s}^{\alpha\beta}
&=
-\frac{\pi e^2}{\hbar}\delta(\Omega)
\sum_{\bm G}
\mathcal C_{1s,-\bm G}^{\alpha}
U_{12}(\bm G)
\mathcal C_{2s,\bm G}^{\beta}
-
\frac{e^2}{\hbar}
\sum_{\bm G}
U_{12}(\bm G)
\Im \pdv{}{\Omega}
\left[
\mathcal D_{1s,-\bm G}^{\alpha}(\Omega)
\mathcal D_{2s,\bm G}^{\beta}(\Omega)
\right]_{\Omega=0}.
\label{eq:drag-full}
\end{align}

The first term represents the intraband contribution, while the second term arises from interband processes. The functions $\mathcal C$ and $\mathcal D$ are defined as
\begin{align}
\mathcal C_{ls,\bm G}^{\alpha}
&=
-\frac{2}{\vol}
\sum_{n,\bm k,\xi}
\Lambda^{(n,n,\xi)}_{ls;\alpha}(\bm k)
\lambda^{(n,n,\xi)}_{l;\bm G}(\bm k)
f'(\epsilon_{l,n,\bm k,\xi}-\mu_l),
\\
\mathcal D_{ls,\bm G}^{\alpha}(\Omega)
&=
\frac{2}{\vol}
\sum_{n\neq m,\bm k,\xi}
\Lambda^{(n,m,\xi)}_{ls;\alpha}(\bm k)
\lambda^{(m,n,\xi)}_{l;\bm G}(\bm k)
\frac{
f(\epsilon_{l,n,\bm k,\xi}-\mu_l)
-
f(\epsilon_{l,m,\bm k,\xi}-\mu_l)
}{
\Omega+\epsilon_{l,n,\bm k,\xi}-\epsilon_{l,m,\bm k,\xi}
},
\end{align}
where $\Lambda^{(n,m,\xi)}_{ls;\alpha}(\bm k) \equiv \Lambda^{(n,m,\xi)}_{ls;\alpha,\bm 0}(\bm k,0)$ and $\lambda^{(n,m,\xi)}_{l;\bm G}(\bm k) \equiv \lambda^{(n,m,\xi)}_{l;\bm G}(\bm k,0)$. We also note that due the orthonormalization of Bloch states in \cref{eq:blochstate} imply $\lambda_{l;\bm 0}^{(n,m,\xi)} = \delta_{n,m}$, so $\mathcal D_{ls,\bm 0}^{\alpha}(\Omega) = 0$. $\mathcal C_{ls,\bm 0}^{\alpha}$ also vanish as shown in the main text. For the interlayer Coulomb interaction $U_{12}(\bm q)$, we consider a single-gated device geometry. Specifically, the graphene–hBN–graphene heterostructure is assumed to be located a distance $d_g$ from a back gate. In this case the interaction takes the form
$U_{12}(\bm q)= e^2 e^{-|\bm q| d}(1-e^{-2|\bm q| d_g})/(2\epsilon_0\epsilon |\bm q|)$,
where $\epsilon$ is the dielectric constant of the surrounding medium. In the calculations we use $\epsilon=5$ and $d_g=10\,\mathrm{nm}$.

In \cref{app:TR} we show that when time-reversal symmetry is present, the charge drag (i.e., \cref{eq:drag-full} with $s=c$) vanishes to linear order. Consequently, for time-reversal-symmetric systems, the leading contribution to charge drag in moiré systems appears only at second order, requiring a generalization of the standard drag formalism \cite{Kamenev1995,Flensberg1995} that incorporates umklapp processes. On the other hand, if time-reversal symmetry is spontaneously broken, e.g., by spin or valley polarization of the ground state, the charge-drag conductivity can remain finite already at linear order.

The situation is different for valley drag. For $s=v$, both the intraband and interband contributions in \cref{eq:drag-full} remain finite even in the presence of time-reversal symmetry. However, the interband term is typically small because the matrix elements $\lambda^{(m,n,\xi)}_{l;\bm G}(\bm k)$ with $n\neq m$ are strongly suppressed. When the chemical potential in both layers lies in a metallic band and a phenomenological disorder broadening $\hbar\delta$ is included, the intraband contribution dominates. The interlayer valley conductivity can then be approximated as
\begin{align}
\sigma_{12,v}^{\alpha\beta}
&\approx
-\frac{e^2}{h}
\frac{2\pi}{\hbar\delta}
\sum_{\bm G}
\mathcal C_{1v,-\bm G}^{\alpha}
U_{12}(\bm G)
\mathcal C_{2v,\bm G}^{\beta},
\label{eq:app:intraband-trm}
\\
\mathcal C_{lv,\bm G}^{\alpha}
&=
-\frac{2}{\vol}
\sum_{n,\bm k,\xi}
\xi\, v^{(n,\xi)}_{l,\alpha}(\bm k)
\lambda^{(n,n,\xi)}_{l;\bm G}(\bm k)
f'(\epsilon_{l,n,\bm k,\xi}-\mu_l),
\end{align}
where we have used the identity $\Lambda^{(n,n,\xi)}_{lv;\alpha}(\bm k)=\xi v^{(n,\xi)}_{l,\alpha}(\bm k)$.
\end{widetext}

\section{Implications of time-reversal symmetry}\label{app:TR}
Let \(\psi_{n\bm\kappa}(\bm r)=e^{i\bm\kappa\cdot\bm r}u_{n\bm\kappa}(\bm r)\) be a Bloch eigenstate. 
We label momenta relative to valley centers \(\bm\kappa=\bm K_\xi+\bm k\) with \(\xi=\pm1\) and \(|\bm k|\ll|\bm K_\xi|\).
The time-reversal operator \(\mathcal T\) is antiunitary. For spinless bands, \(\mathcal T^2=+1\) and
\begin{align*}
	\mathcal T \psi_{n\bm\kappa}(\bm r)=\psi_{n,-\bm\kappa}(\bm r)
	\quad\Rightarrow\quad
	\mathcal T u_{n\bm\kappa}(\bm r)=u^{*}_{n,-\bm\kappa}(\bm r).
\end{align*}
Since \(-\bm K_\xi=\bm K_{-\xi}+\bm G_0\) (some reciprocal vector \(\bm G_0\)), we may choose a smooth gauge so that
\begin{align*}
	\mathcal T:\ (\xi,\bm k)\mapsto(-\xi,-\bm k),\qquad
	u_{n,\bm k,\xi}(\bm r)\mapsto u^{*}_{n,-\bm k,-\xi}(\bm r).
\end{align*}
In the plane–wave basis, \(u_{n,\bm k,\xi}(\bm r)=\sum_{\bm G}u_{n,\bm k,\xi}^{(\xi)}(\bm G) e^{i\bm G\cdot\bm r}\), this implies
\begin{align*}
	u_{n,-\bm k,\xi}(\bm G)=u^{*}_{n,\bm k,\xi}(-\bm G).
\end{align*}
Time reversal acts as \(\mathcal T H_l^{(\xi)}(\bm k)\mathcal T^{-1}=H_l^{(-\xi)}(-\bm k)\). So,
\begin{align*}
	\mathcal T \Gamma^{(\xi)}_{l;\alpha}(\bm k) \mathcal T^{-1}
	=\frac{1}{\hbar} \partial_{k_\alpha}\big[H^{(-\xi)}(-\bm k)\big]
	=-\Gamma^{(-\xi)}_{l;\alpha}(-\bm k).
\end{align*}
From the above and using \cref{eq:form-fac-i,eq:form-fac-ii} we arrive at
\begin{align*}
	&\lambda^{(n,m,\xi)}_{l,\bm G}(\bm k) = \big[\lambda^{(m,n,-\xi)}_{l,-\bm G}(-\bm k)\big]^{*},\nonumber\\
	&\Lambda^{(n,m,\xi)}_{l,c,\alpha}(\bm k) = -\big[\Lambda^{(m,n,-\xi)}_{l,c,\alpha}(-\bm k)\big]^{*}.
\end{align*}
Using these, it is easy to see that charge drag in the first order [\cref{eq:drag-full} with $s=c$] vanishes. 
\section{Zero temperature limit of the intraband valley-drag susceptibility}\label{app:chi-deriv}

Starting from the intraband valley susceptibility,
\begin{align}
\mathcal C^{\alpha}_{lv,\bm G}
	&= -\frac{2}{\vol}\sum_{n,\bm k,\xi}
\xi\, v^{(n,\xi)}_{l,\alpha}(\bm k)\,
	\lambda^{(n,n,\xi)}_{l,\bm G}(\bm k)\;
	f'(\varepsilon_{l,n,\bm k,\xi}-\mu_l),
	\label{eq:chi-intra-val-app}
\end{align}
where $v^{(n,\xi)}_{l,\alpha}(\bm k) = \frac{1}{\hbar}\partial_{k_\alpha}\varepsilon_{l,n,\bm k,\xi}$ is the band velocity. Substituting this, we write
\begin{align}
\mathcal C^{\alpha}_{lv,\bm G}
	&= -\frac{2}{\hbar\vol}\sum_{n,\bm k,\xi}
\xi\, \partial_{k_\alpha}\varepsilon_{l,n,\bm k,\xi}\;
	\lambda^{(n,n,\xi)}_{l,\bm G}(\bm k)\;
	f'(\varepsilon_{l,n,\bm k,\xi}-\mu_l).
\end{align}
Using the identity $\partial_{k_\alpha}\varepsilon \cdot f'(\varepsilon - \mu) = \partial_{k_\alpha} f(\varepsilon - \mu)$, we obtain
\begin{align}
\mathcal C^{\alpha}_{lv,\bm G}
	&= -\frac{2}{\hbar\vol}\sum_{n,\bm k,\xi}
\xi\, \big[\partial_{k_\alpha} f(\varepsilon_{l,n\bm,k,\xi}-\mu_l)\big]\;
    \lambda^{(n,n,\xi)}_{l,\bm G}(\bm k).
\end{align}
Integrating by parts over the Brillouin zone (the boundary terms vanish by periodicity),
\begin{align}
\mathcal C^{\alpha}_{lv,\bm G}
	&= \frac{2}{\hbar\vol}\sum_{n,\bm k,\xi}
\xi\, f(\varepsilon_{l,n\bm k}^{(\xi)}-\mu_l)\;
	\partial_{k_\alpha}
    \lambda^{(n,n,\xi)}_{l,\bm G}(\bm k).
    \label{eq:chi-ibp}
\end{align}
To obtain the zero-temperature limit, we return to \cref{eq:chi-intra-val-app} and use $f'(\varepsilon - \mu) \to -\delta(\varepsilon - \mu)$ as $T\to 0$. Converting the momentum sum to an integral,
\begin{align}
\mathcal C^{\alpha}_{lv,\bm G}
	&= \frac{2}{(2\pi)^2}\sum_{n,\xi}
    \int_{\mathrm{BZ}} d^2\bm k\;
\xi\, \nonumber \\
&\times [v^{(n,\xi)}_{l,\alpha}(\bm k)\,
    \lambda^{(n,n,\xi)}_{l,\bm G}(\bm k)
	\delta(\varepsilon_{l,n,\bm k,\xi}-\mu_l)].
\end{align}
The delta function restricts the integral to the Fermi surface $\mathrm{FS}_{l,n,\xi}$, defined by $\varepsilon_{l,n,\bm k,\xi} = \mu_l$. Decomposing $d^2\bm k$ into components along and perpendicular to the Fermi surface, $d^2\bm k = d\ell_{\bm k}\, dk_\perp$, and using $\delta(\varepsilon - \mu) = \delta(k_\perp)/|\nabla_{\bm k}\varepsilon| = \delta(k_\perp)/({\hbar |{\bm v}_{l,\alpha}^{(n,\xi)}(\bm k)|})$, we arrive at
\begin{align}
\mathcal C^{\alpha}_{lv,\bm G}\Big|_{T=0}
	= \frac{2}{(2\pi)^2}\sum_{n,\xi}\oint_{\mathrm{FS}_{l,n,\xi}}
	\frac{\xi\, d\ell_{\bm k}}{|{\bm v}_{l,n}^{(\xi)}(\bm k)|}\; v^{(\xi,\alpha)}_{l,n}(\bm k)\;
    \lambda^{(n,n,\xi)}_{l,\bm G}(\bm k).
\end{align}

\section{Sommerfeld expansion of the intraband susceptibility}\label{app:sommerfeld}

We define the spectral function
\begin{align}
	\mathcal{F}^{\alpha}_{l,\bm G}(\varepsilon) = \frac{1}{\vol}\sum_{n,\bm k,\xi} \xi\, v^{(\xi,\alpha)}_{l,n}(\bm k)\,
    \lambda^{(n,n,\xi)}_{l,\bm G}(\bm k)
    \delta(\varepsilon - \varepsilon_{l,n,\bm k,\xi}),
    \label{eq:spectral-fn}
\end{align}
so that \cref{eq:chi-intra-val-app} may be rewritten as
\begin{align}
	\mathcal C^{\alpha}_{lv,\bm G}(T) = -2\int_{-\infty}^{\infty} d\varepsilon\; \mathcal{F}^{\alpha}_{l,\bm G}(\varepsilon)\, f'(\varepsilon - \mu_l).
	\label{eq:chi-energy}
\end{align}
Note that $\mathcal{F}$ is temperature-independent; all temperature dependence enters through $f'$.
At zero temperature, $f'(\varepsilon - \mu) \to -\delta(\varepsilon - \mu)$, yielding
\begin{align}
	\mathcal C^{\alpha}_{lv,\bm G} = 2\,\mathcal{F}^{\alpha}_{l,\bm G}(\mu_l).
	\label{eq:chi-zero-T}
\end{align}

For finite but low temperatures ($k_B T \ll E_F$), we apply the standard Sommerfeld expansion. For a smooth function $g(\varepsilon)$,
\begin{align}
	\int_{-\infty}^{\infty} d\varepsilon\; g(\varepsilon)\,f'(\varepsilon - \mu) = -g(\mu) - \frac{\pi^2}{6}(k_BT)^2\, g''(\mu) + \mathcal{O}(T^4).
    \label{eq:sommerfeld-identity}
\end{align}
Applying this to \cref{eq:chi-energy} with $g = \mathcal{F}^{\alpha}_{l,\bm G}$,
\begin{align}
	\mathcal C^{\alpha}_{lv,\bm G} &= 2\,\mathcal{F}^{\alpha}_{l,\bm G}(\mu_l) + \frac{\pi^2}{3}(k_BT)^2\, \frac{\partial^2 \mathcal{F}^{\alpha}_{l,\bm G}}{\partial \varepsilon^2}\bigg|_{\varepsilon = \mu_l} + \mathcal{O}(T^4).
    \label{eq:chi-sommerfeld-F}
\end{align}
Using \cref{eq:chi-zero-T}, $\mathcal{F}^{\alpha}_{l,\bm G}(\mu_l) = \frac{1}{2}\mathcal C^{\alpha}_{lv,\bm G}(0)$, and therefore $\frac{\partial^2 \mathcal{F}}{\partial \varepsilon^2}\big|_{\mu} = \frac{1}{2}\frac{\partial^2 \mathcal C(0)}{\partial \mu_l^2}$, where the derivative is taken with respect to $\mu_l$ at fixed $T=0$. Substituting,
\begin{align}
	\mathcal C^{\alpha}_{lv,\bm G}(T) &= \mathcal C^{\alpha}_{lv,\bm G}(0) + \frac{\pi^2}{6}(k_BT)^2\, \frac{\partial^2 \mathcal C^{\alpha}_{lv,\bm G}(0)}{\partial \mu_l^2} + \mathcal{O}(T^4).
    \label{eq:chi-sommerfeld}
\end{align}

Since the valley drag conductivity in \cref{eq:netsigma} is bilinear in the susceptibilities $\mathcal C$ of the two layers, substituting \cref{eq:chi-sommerfeld} for each layer yields
\begin{align}
	\sigma_{12,v}^{\alpha \beta}(T) = \sigma_{12,v}^{\alpha \beta}(0) + A^{\alpha \beta}_{v}\, T^2 + \mathcal{O}(T^4),
\end{align}
where $\sigma_{12,v}^{\alpha \beta}(0)$ is the zero-temperature valley drag conductivity and the coefficient $A^{\alpha \beta}_{v}$ involves both the zero-temperature susceptibilities and their second derivatives with respect to the chemical potentials $\mu_1$, $\mu_2$.

\section{Phenomenological broadening of reciprocal-lattice contributions}

In \cref{eq:app:intraband-trm}, the drag conductivity is given by the sum over all discrete reciprocal lattice vectors $\bm G =n \bm G_1 + m \bm G_2$. Introducing a Dirac delta allows us to recast this discrete sum as an integral over a continuous reciprocal-space variable $\mathbf{g}$,
\begin{align}
	\sigma_{12,v}^{\alpha\beta} &\approx -\frac{e^2}{h} \frac{2 \pi}{\hbar \delta} \int_{\mathbb R^2} \dd \bm g   
	\mathcal{C}_{1v, - \bm g}^{\alpha} U_{12}(\bm g) \mathcal{C}_{2v,\bm g}^{\beta}
    \sum_{\bm G} \delta(\bm g - \bm G)
\end{align}
The integration runs on the infinite 2D plane. This representation is convenient for incorporating lattice imperfections, which effectively broaden the reciprocal-lattice points.

To model this effect phenomenologically, we replace each Dirac delta by a Gaussian peak of width $\eta$,
\begin{widetext}
\begin{align}
\tilde \sigma_{12,v}^{\alpha\beta}
= -\frac{e^2}{h} \frac{2 \pi}{\hbar \delta} \int_{\mathbb R^2} \dd \bm g   
	\mathcal{C}_{1v, - \bm g}^{\alpha} U_{12}(\bm g) \mathcal{C}_{2v,\bm g}^{\beta}
    \sum_{\bm G} 
    \frac{1}{2\pi \eta^2}
    \exp\!\left[-\frac{\|\bm g - \bm G\|^2}{2\eta^2}\right].
    \label{eq:app:gaussian}
\end{align}
This replaces each reciprocal-lattice point by a gaussian broadened regions, mimicking relaxation-induced disorder.

For compactness, we define \( \Xi^{\alpha \beta}_{12,v}(\bm g) = \mathcal{C}_{1v, - \bm g}^{\alpha} U_{12}(\bm g) \mathcal{C}_{2v,\bm g}^{\beta}\) and for small $\eta$, we expand this function about each reciprocal lattice point $\bm{G}$,
\begin{align}
 \Xi^{\alpha \beta}_{12,v}(\bm g)
  =  \Xi^{\alpha \beta}_{12,v}(\bm G)
  + (\bm g - \bm G) \cdot \nabla \Xi^{\alpha \beta}_{12,v}(\bm G) 
  + \frac{1}{2} (\bm g - \bm G)^T 
    \mathcal H(\bm G) 
    (\bm g - \bm G)
  + \dots ,
\end{align}
\end{widetext}
where $\mathcal H(\mathbf{G}_{nm})$ is the Hessian of $\Xi^{\alpha \beta}_{12,v}$.
Putting this back in \cref{eq:app:gaussian} and 
integrating term by term, we find that the linear term vanishes by symmetry, while the quadratic term yields $\eta^2 \,\mathrm{Tr}[\mathcal H(\mathbf{G}_{nm})]$. This leads to the expansion
\begin{align}
\tilde \sigma_{12,v}^{\alpha\beta} 
= \sigma_{12,v}^{\alpha\beta} 
- \frac{\eta^2}{2} \frac{e^2}{h} \frac{2\pi}{\hbar \delta} \sum_{\bm G} \nabla^2 \Xi(\bm G) 
+ O(\eta^4).
\end{align}
with $\nabla^2 = \partial_{G_x}^2 + \partial_{G_y}^2$. The leading term reproduces the original discrete expression for $\sigma_{12,v}^{\alpha\beta}$, while the first correction is controlled by the local curvature of $\Xi$ at each reciprocal-lattice point.

\bibliography{ref3.bib}
\end{document}